      [1999/12/01 v1.4c Il Nuovo Cimento]
\documentclass{cimento}
\usepackage{amssymb}
\usepackage{epsfig}
\newcommand\cm{\,{\rm cm}}
\newcommand\G{\,{\rm G}}

\def\d{{\rm d}}
\def\ne{n_{\rm e}}

\def\taut{\tau_{\rm T}}

\def\sigmat{\sigma_{\rm T}}
\def\lamb{\Lambda_{\rm B}}
\def\lams{\Lambda_{\rm T}}
\def\lamsyn{\Lambda_{\rm s}}
\def\lamcomp{\Lambda_{\rm c}}

\def\gammab{\gamma_{\rm b}}
\def\gammam{\langle\gamma\rangle}
\def\Bcr{B_{\sf QED}}

\title{Gamma-ray burst spectra from continuously accelerated electrons}
\author{Juri~Poutanen\from{ins:x} \thanks{NORDITA corresponding fellow.} \atque
Boris~E.~Stern\from{ins:x} \from{ins:a} \from{ins:b}
}
\instlist{\inst{ins:x} Astronomy Division, P.O.Box 3000, 90014 University of Oulu,
Finland
  \inst{ins:a} Institute for Nuclear Research, Russian Academy of Sciences,
 Moscow 117312, Russia
 \inst{ins:b} Astro Space Center of Lebedev Physical Institute,
Profsoyuznaya 84/32, Moscow 117997, Russia
}
\PACSes{\PACSit{98.70.Rz}{$\gamma$-ray sources; $\gamma$-ray bursts} }
\begin{document}

\maketitle

\begin{abstract}
We discuss here   constraints on the particle acceleration
models from the observed  gamma-ray bursts spectra.
The standard synchrotron shock model assumes that
some fraction of available energy is given instantaneously to the electrons
which are injected at high Lorentz factor.
The emitted spectrum in that case corresponds to the   spectrum of
cooling electrons, $F_{\nu}\propto \nu^{-1/2}$,
  is much too soft to
account for the majority of the observed spectral slopes.
We show that continuous heating of electrons over the life-time of a
source  is needed to produce hard observed spectra.
In this model, a prominent peak develops in the electron
distribution at energy which is a strong function of Thomson
optical depth $\taut$ of heated electrons (pairs).
At $\taut\gtrsim 1$, a typical electron Lorentz factor
$\gammam\sim 1-2$ and
quasi-thermal Comptonization operates.
It produces spectrum peaking at a too high energy.
Optical depths below $10^{-4}$ would be difficult to imagine
in any physical scenario.
At $\taut\approx10^{-4}$--$10^{-2}$, $\langle\gamma\rangle\sim 30-100$ and
synchrotron self-Compton radiation
is the main emission mechanism.
 The synchrotron peak should be observed at 10--100 eV, while the
self-absorbed low-energy tail with  $F_{\nu}\propto \nu^{2}$ can
produce the prompt optical emission (like in the case of
GRB~990123). The first Compton scattering radiation by nearly
monoenergetic electrons peaks in the BATSE energy band and
can   be as hard as $F_{\nu}\propto
\nu^{1}$ reproducing the hardness of most of the observed GRB
spectra. The second Compton peak should be observed in the
high-energy gamma-ray band, possibly being responsible for the
10--100 MeV emission detected in GRB 941017. A significant
electron-positron pair production reduces the available energy per
particle, moving   spectral peaks to lower energies as the burst
progresses.
\end{abstract}

\section{Introduction}

The time-resolved gamma-ray burst (GRB) spectra have the mean
observed photon spectral index $\alpha$ close to $-1$ (i.e. $F_\nu\propto \nu^{0}$)
and some spectra can be as hard as $F_\nu\propto~\nu^1$ \cite{pr00}.
Synchrotron shock models (see, e.g., \cite{p99})
assume that a fraction of the available
energy is given instantaneously to the electrons which are injected
at Lorentz factor $\gamma\gtrsim 10^3$. The typical cooling time-scale
is smaller that the light crossing time of the source by a factor
$\gamma\ell$, where  the compactness $\ell$ is about $0.1-10$
for typical GRB parameters \cite{sp04}. Because the cooling time is
small, one observes the time-averaged spectrum of cooling electrons
\cite{gc99,gcl00}.

For relativistic electrons, the photon emission frequency  is
$\nu\propto \gamma^m$, where $m=2$
for synchrotron or synchrotron-self-Compton (SSC) radiation.
Thus, the rate of the frequency change
$\d\nu/\d t \propto \gamma^{m-1} \d\gamma/\d t \propto \gamma^{m-1} P
\propto \nu^{(m-1)/m} P$,
where $P=\d E/\d t$ is the emitted power. The time-averaged flux is then
\begin{equation}
F_{\nu}=\frac{\d E}{\d \nu}=\frac{\d E}{\d t} \frac{\d t}{\d \nu}=
\frac{P}{\d \nu/\d t}\propto \nu^{-(m-1)/m} ,
\end{equation}
and the ``cooling'' spectrum is
$F_{\nu}\propto \nu^{-1/2}$ (i.e. $\alpha=-3/2$)
for synchrotron  as well as
for SSC \cite{gcl00}, which is much   softer that those
observed from GRBs.
The immediate conclusion from this simple exercise is that
{\it all} the models involving injection of electrons at
large energies should be rejected.

The efficient cooling can be prevented if the energy
is supplied to the electrons in small steps.
In this paper, we discuss consequences of the assumption that
the available energy is supplied continuously to the particles over the
life-time of a source. We study how the resulting spectrum
depends on main parameters and determine  parameter
range where model spectra are consistent with the majority
of the observed ones.

\section{Spectra from continuously heated electrons}

\subsection{Model setup}

A number of physical models can produce a source of continuously
heated particles. These can be plasma instabilities behind ultrarelativistic
shocks \cite{h04}
or dissipation of the magnetic energy in a Poynting flux dominated
outflow (e.g. \cite{gs05}).
Therefore, we consider a toy model where energy is injected to the emission
region with the constant rate during comoving time $R'/c$. We
adopt a slab geometry with the thickness $\Delta=0.1$ (in units of $R'$)
which may depend  on the specific scenario.
The energy is injected uniformly over the slab volume
in a form of acceleration of electrons (and pairs)
which  obtain equal amount of energy per unit time.
The model is fully described by four parameters: (i) the initial
Thomson optical depth across the slab, $\tau_0=\ne\sigmat \Delta\
R'$; (ii) the comoving size $R'$; (iii) the dissipation
compactness, $\ell=E_{\rm rad} \sigmat / (m_e c^2 \Gamma^3 4 \pi R'^2)$
related to the observed isotropic energy release $E_{\rm rad}$ and the
ejecta bulk Lorentz factor $\Gamma$,
which determines the rate of pair production;
and (iv) the magnetic compactness
$\lamb=B^2 R' \sigmat /(8\pi m_e c^2)$
determining the role of synchrotron radiation.
See \cite{sp04} for details.

\subsection{Radiative processes}

Let us first consider how particles (electron and positrons) of
Lorentz factor $\gamma$ are heated and how do they cool. The energy
gain rate of a particle is simply given by the heating rate
$\propto \ell$ divided by the number of particles (which is
proportional to the total Thomson optical depth, including pairs,
across the slab, $\taut$). Particles cool by emitting
synchrotron radiation and by scattering this radiation
(SSC). The energy balance equation can be
written as:
\begin{equation}
\label{eq:enebal} \frac{\d \gamma}{\d t'}= \frac{\ell}{\taut} -
\frac{4}{3}(\eta\lamb + \lams)\gamma^2 ,
\end{equation}
where  $\eta<1$ accounts for the reduced synchrotron cooling due to
synchrotron self-absorption and $\lams$ is the compactness
corresponding to the energy density of soft photons in the Thomson
regime.  It
 can be expressed as a sum of the synchrotron $\lamsyn=y
\eta\lamb$ and first Compton scattering $\lamcomp=y \eta\lamsyn$
energy densities, where $y\approx \taut \langle \gamma^2\rangle$
is the Compton parameter.
If mean Lorentz factor $\gammam \lesssim 10$, then
one need to account for further scattering orders.
The typical cooling time, $t_{\rm cool}  \sim
(R'/c)/[(\lams + \eta\lamb) \gamma ]$,
is orders of magnitude smaller than the light-crossing time $R'/c$
for GRB conditions.

The balance between heating and cooling is achieved at
\begin{equation}
\label{eq:gampeak} \gammab\approx \sqrt{\ell/(\lams+\eta\lamb)}
\taut^{-1/2} \approx {\rm a\ few}\times \taut^{-1/2} ,
\end{equation}
and thus $y={\rm a\ few}$.
Particles with
$\gamma>\gammab$ lose energy faster than they gain it, while at
$\gamma<\gammab$ the situation is opposite resulting in a
narrow distribution peaked at $\gammab$.

Main radiative process is determined by
the instantaneous optical depth $\taut$, which can be
significantly larger than the initial $\tau_0$ because
of pair production at $\ell \gtrsim 0.1$.
If $\taut \lesssim 10^{-8}$, then $\gammab\gtrsim 10^4$ and
the main emission mechanism is synchrotron.
A typical  energy of a synchrotron photon in
the frame comoving with the ejecta
(in units $m_e c^2$) is $\epsilon_{\rm s}\approx b \gammab^2\sim 1$,
for $b\sim 10^{-8}$ (where $b=B/\Bcr$ and $\Bcr=4.4\times10^{13}\G$)
and Compton scattering is in Klein-Nishina limit.

If $\taut\sim10^{-4}$, $\gammam\sim 10^2$ and
$\epsilon_{\rm s}\sim 10^{-4}$. This time, first Compton
scattering peak is at $\epsilon_{\rm c1}\sim 1$, and one
can neglect further scatterings. At $\taut\sim10^{-2}-10^{-3}$,
second Compton scattering will dominate the energy output.
For high compactness source $\taut$ can reach 1,
then $\gammam\sim 1-2$ and multiply Compton scatterings
(quasi-thermal Comptonization) have to be accounted for.

\subsection{Simulations}
\label{sec:results}

The simulations were performed using a Large Particle Monte Carlo
code described in \cite{st95}. It treats
Compton scattering, photon-photon pair production and  pair
annihilation, synchrotron radiation and synchrotron
self-absorption. The electron/pair and photon distributions are computed
self-consistently.

\begin{figure}
\centerline{\epsfig{file=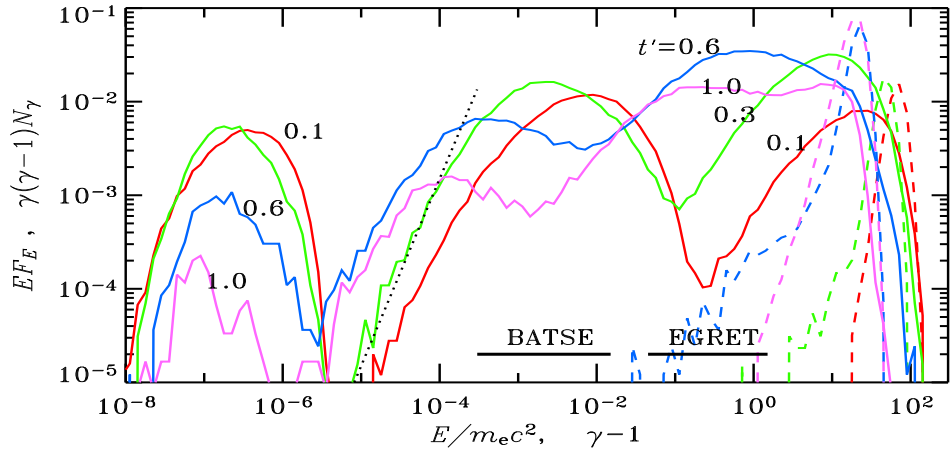,width=13cm,height=6.0cm}}
\caption{
Instantaneous  (comoving frame)  photon spectra
(solid curves) and corresponding electron
distributions (dashed)
at times of $0.1, 0.3$, $0.6$, and $1$ (in units $R'/c$).
The spectra consist of a low-energy  synchrotron
 and two Compton scattering orders.
The hardest possible power-law  $F_\nu\propto \nu^1$
reachable at the low-energy slope of the first Compton bump
is shown with dotted line. The BATSE
(20--1000 keV) and EGRET (3--100 MeV) bands (redshifted to a comoving frame)
are marked.} \label{fig:spectra}
\end{figure}

As an example, we consider  $R'=10^{13}\cm$,
$\tau_0=6\times 10^{-4}$, $\ell=3$
(corresponding to $\Gamma\approx 130$ and $E_{\rm rad}=10^{52}$ erg), and
$\lamb=0.3$, which are typical for GRB ejecta \cite{sp04}.
 The resulting
evolution of broad-band photon spectra and
electron distributions is shown in Fig.~\ref{fig:spectra}.
At the start of simulations $\taut=\tau_0$ and  $\gammam\sim80$
(see dashed curves). Partially self-absorbed synchrotron
peaks at $\epsilon_{\rm s}\sim 3\times 10^{-7}$ (UV in the observer
frame)
with low-energy tail $F_\nu\propto \nu^2$ may be
responsible for the prompt optical emission seen in GRB 990123 \cite{a99}.
When $\taut$ starts to grow due to  pair production,
$\gammam$ drops as  $1/\sqrt{\tau}$, and
the first Compton peak moves to
softer energies crossing the `BATSE window'. This
spectral evolution is  consistent with the observed in
time-resolved pulses \cite{ford95,rs02}.
The low-energy slope of this component,
corresponding to Compton scattering by
isotropic monoenergetic electrons,
can as hard as $F_\nu\propto\nu^1$ (i.e. $\alpha=0$),
reproducing most of the observed GRB spectra.
The second Compton component peaking at $\sim 10$--$100$ MeV rises
later and decays  on a longer time-scale. It can be responsible
for the delayed  emission observed by EGRET
from GRB 941017 \cite{gon03}. Thus, at $\taut \sim 10^{-2}-10^{-4}$,
the model spectra
reproduce well the observed ones from GRBs, including their hardness,
spectral evolution, as well as optical and 100 MeV  emission.

%%%%%%%%%%%%%%%%%%%%%%%%%%%%%%%%%%%%%%%%%%%%%

Small $\taut$ required for  synchrotron to be responsible
for the BATSE GRBs are difficult to get in any physical scenario.
Additionally, the synchrotron
low energy spectrum $F_\nu\propto\nu^{1/3}$ is softer that the
observed ones \cite{pr00}.
At $\taut\gtrsim1$, the resulting quasi-thermal
Comptonization spectrum may be too soft for most of the GRB spectra.
The emission  peaks at 10--50 keV in the comoving frame of the
ejecta and, for  $\Gamma\sim100$, this peak shifts
 to an uncomfortably high energy \cite{gc99,st99,sp04}.
In conclusion, SSC from continuously heated electrons (pairs)
seems to be the most promising model to explain the
observed properties of GRBs.

\acknowledgments
This research has been supported by the RFBR grant 04-02-16987,
Academy of Finland, Wihuri Foundation,
V\"ais\"al\"a Foundation, and the NORDITA
Nordic project in High Energy Astrophysics.

\end{document}